\documentclass{iau} 
\usepackage{graphicx}
\usepackage{hyperref}

\title[CubeSpec space mission] %% give here short title %%
{The CubeSpec space mission: Asteroseismology of massive stars from time-series optical spectroscopy}

\author[D. M. Bowman et al.]   %% give here short author list %%
{D. M. Bowman$^{1}$, B. Vandenbussche$^1$, H. Sana$^1$, A. Tkachenko$^1$, \\ G. Raskin$^1$, T. Delabie$^2$, B. Vandoren$^2$, P. Royer$^1$, S. Garcia$^1$, \\ T. Van Reeth$^1$, and the CubeSpec Collaboration }

\affiliation{
$^1$Institute of Astronomy, KU Leuven, Celestijnenlaan 200D, 3001 Leuven, Belgium \\ email: {\tt dominic.bowman@kuleuven.be} \\
$^2$Arcsec NV, Blijde Inkomststraat 22, 3000 Leuven, Belgium \\
}

\pubyear{2022}
\volume{361}  %% insert here IAU Symposium No.
\jname{Massive Stars Near and Far}
\editors{N. St-Louis, J. S. Vink \& J. Mackey, eds.}
\begin{document}

\maketitle

\begin{abstract}
The ESA/KU Leuven CubeSpec mission is specifically designed to provide low-cost space-based high-resolution optical spectroscopy. Here we highlight the science requirements and capabilities of CubeSpec. The primary science goal is to perform pulsation mode identification from spectroscopic line profile variability and empower asteroseismology of massive stars. 


\end{abstract}

\firstsection % if your document starts with a section, remove some space above using this command.

\section{Scientific Context}

There is a need to constrain the interior rotation, mixing and angular momentum transport mechanisms of massive stars, as they strongly influence stellar evolution. Asteroseismology -- the study of stellar structure from pulsations -- is an excellent method for probing the interiors of massive stars \cite[(Bowman 2020)]{Bowman2020}. The ongoing NASA TESS mission is providing high-precision light curves of thousands of massive stars \cite[(Ricker et al. 2015)]{Ricker2015}. Complementary time-series spectroscopy is highly advantageous in fully characterising pulsations: spectral line profile variability reveals the geometry of pulsation modes in terms of spherical harmonics \cite[(e.g. Aerts et al. 1994; Briquet et al. 2009)]{Aerts1994, Briquet2003}. The combination of TESS photometry and CubeSpec spectroscopy will unlock the physics of massive star interiors through forward asteroseismic modelling.

\section{CubeSpec Specifications}

\begin{figure}[h]
\begin{center}
\includegraphics[height=4.5cm]{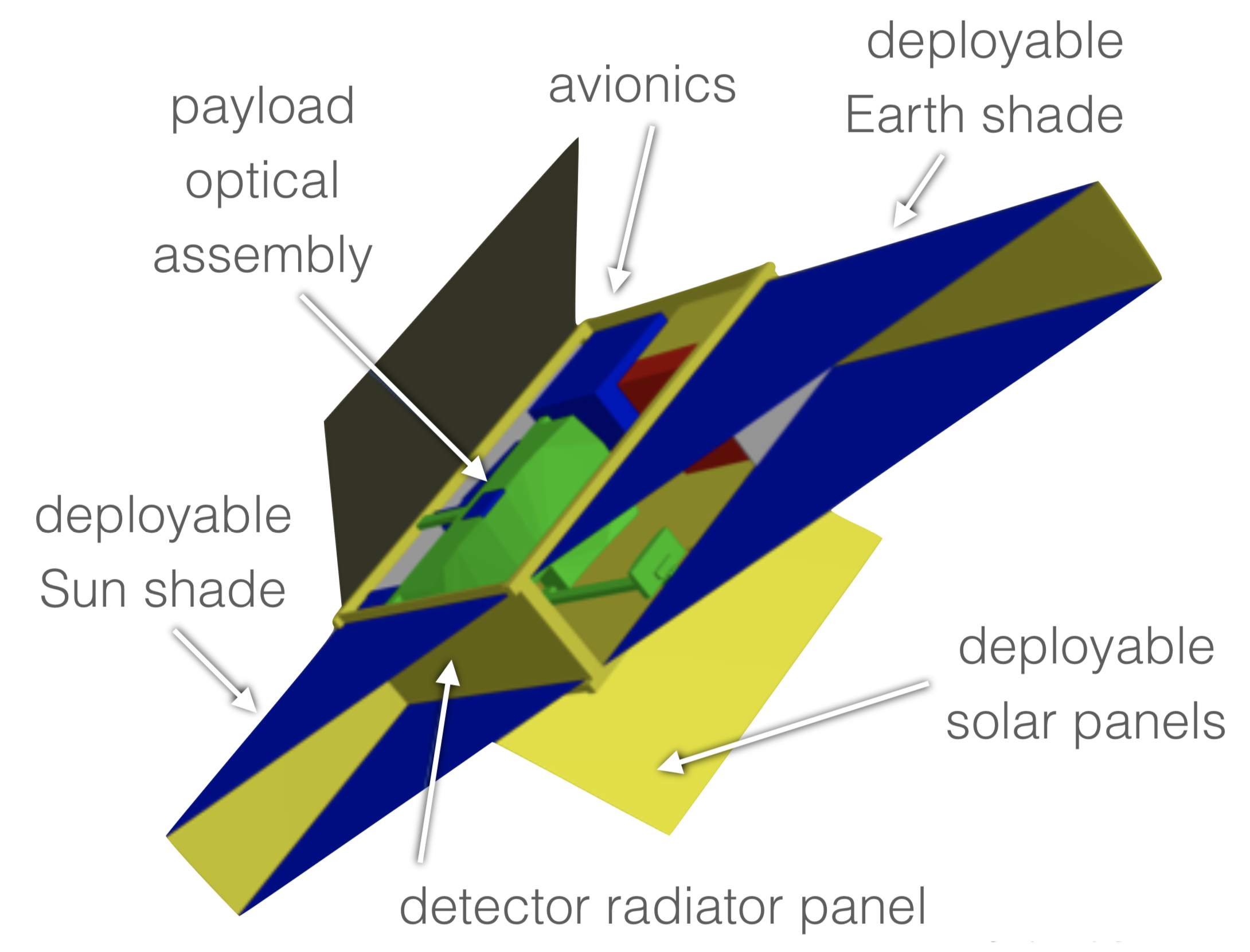} 
\hspace{1cm}
\includegraphics[height=4.5cm]{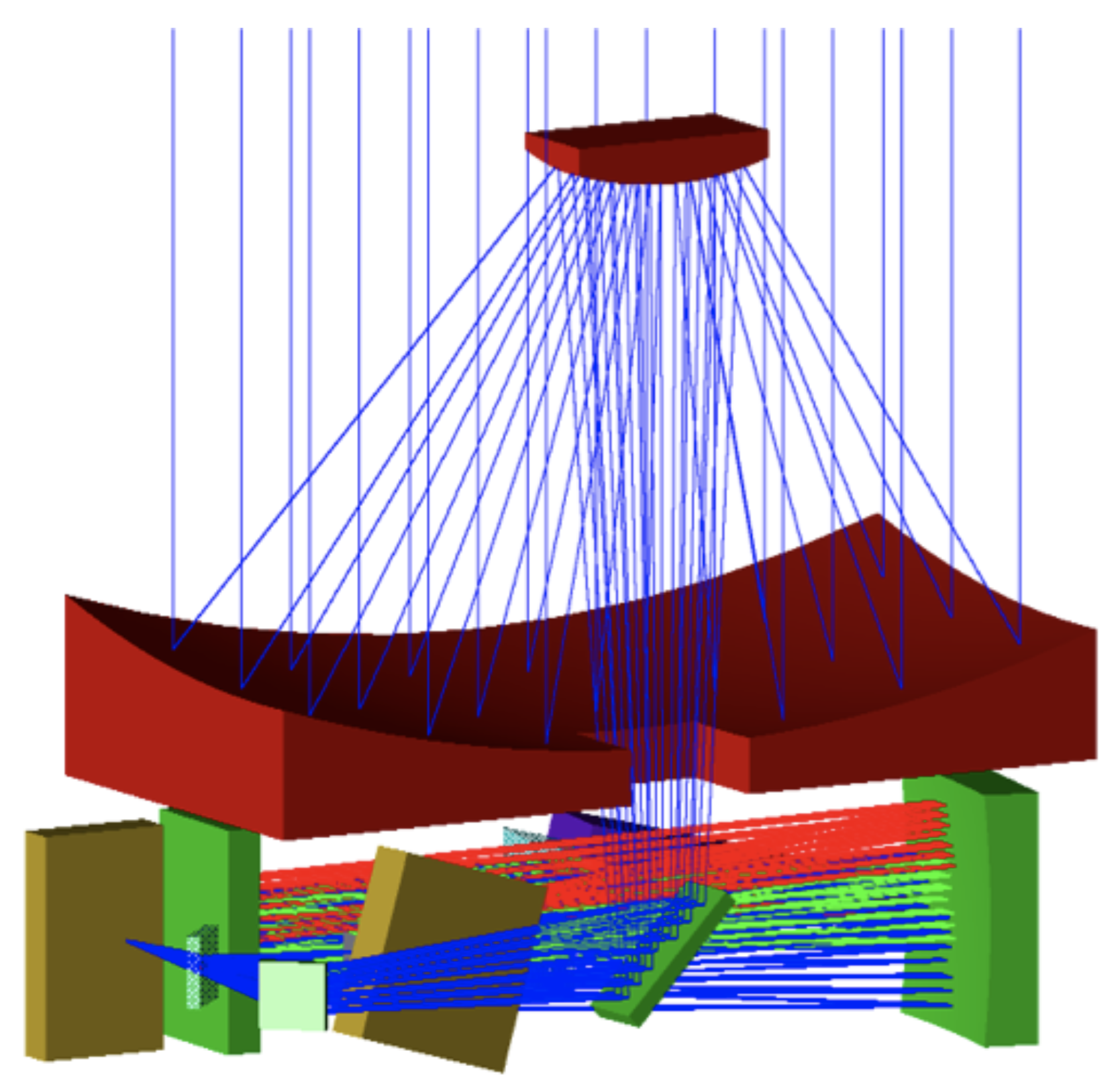} 
\caption{Platform layout (left) and optical design (right) of the CubeSpec space mission.}
\label{figure: design}
\end{center}
\end{figure}

CubeSpec is a 6U cubesat on schedule to be launched in 2024. The platform includes a Cassegrain telescope with a rectangular primary mirror of $9 \times 19$~ cm$^2$, and a compact high-resolution echelle optical spectrograph with a resolving power of $R = 55000$ \cite[(Raskin et al. 2018)]{Raskin2018}. Schematics of the platform and optical design are shown in Fig.~\ref{figure: design}, and on the CubeSpec project website: \href{https://fys.kuleuven.be/ster/instruments/cubespec}{\url{https://fys.kuleuven.be/ster/instruments/cubespec}}.

To achieve the primary science goal to assemble time-series spectroscopy for massive-star asteroseismology, CubeSpec has the following science requirements: (i) High resolving power ($R > 50000$) to resolve perturbations to narrow spectral lines in slowly rotating stars; (ii) High signal-to-noise ratio (SNR~$> 200$) spectra to detect small perturbations relative to the median line profile; (iii) Observing cadence of order 90~min to effectively sample the pulsation periods of order several hours; (iv) Short exposure times of order minutes to avoid pulsational smearing of spectral lines; (v) Mission duration exceeding three months to provide a high frequency precision; and (vi) Optical wavelength range to include the silicon triplet at $\lambda\lambda$ 4552, 4567, and 4574 \AA, which is sensitive to pulsations.

\section{Prioritised Target List}

\begin{figure}[h]
\begin{center}
\includegraphics[height=4.5cm]{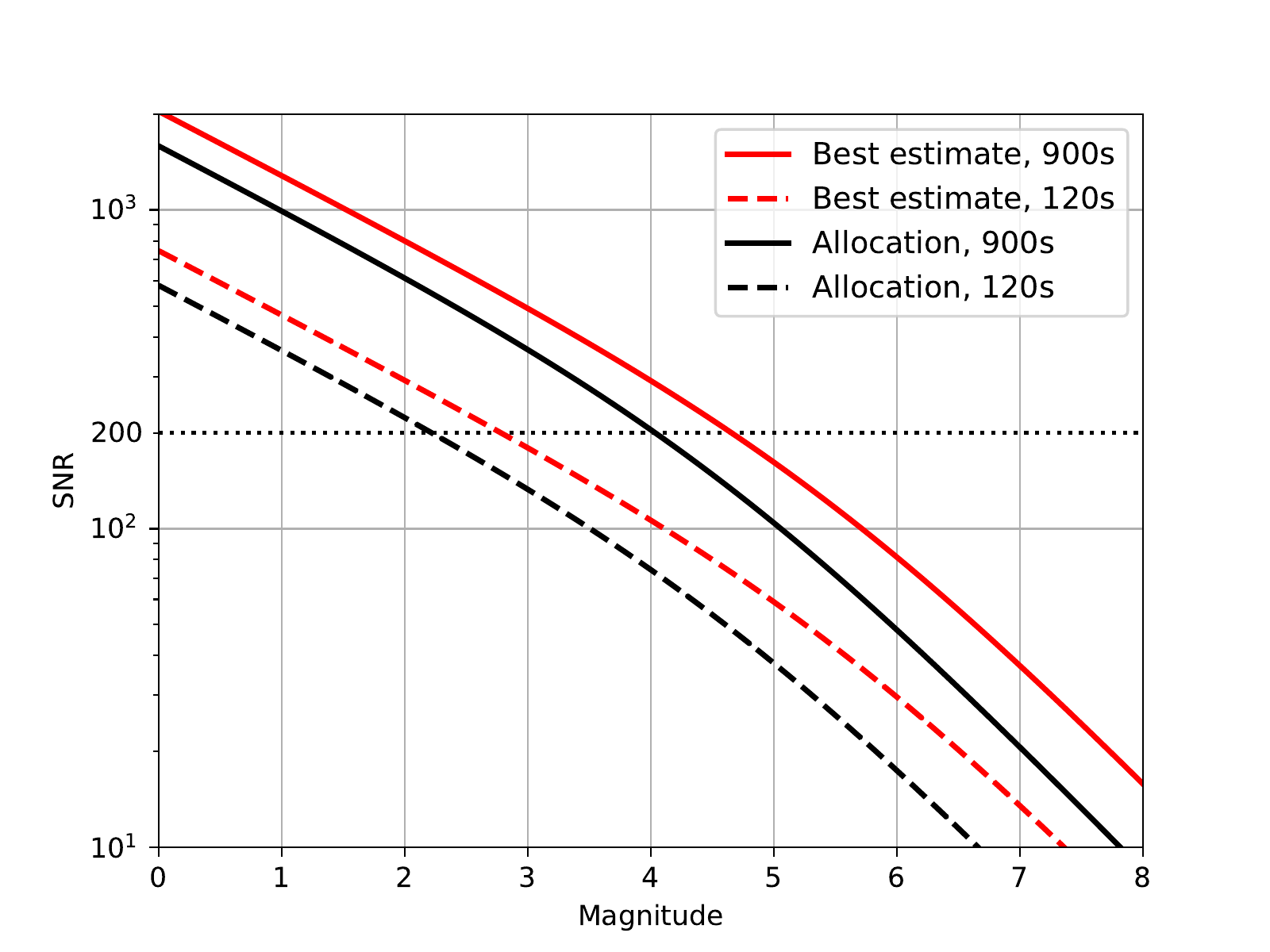} 
\includegraphics[height=4.3cm]{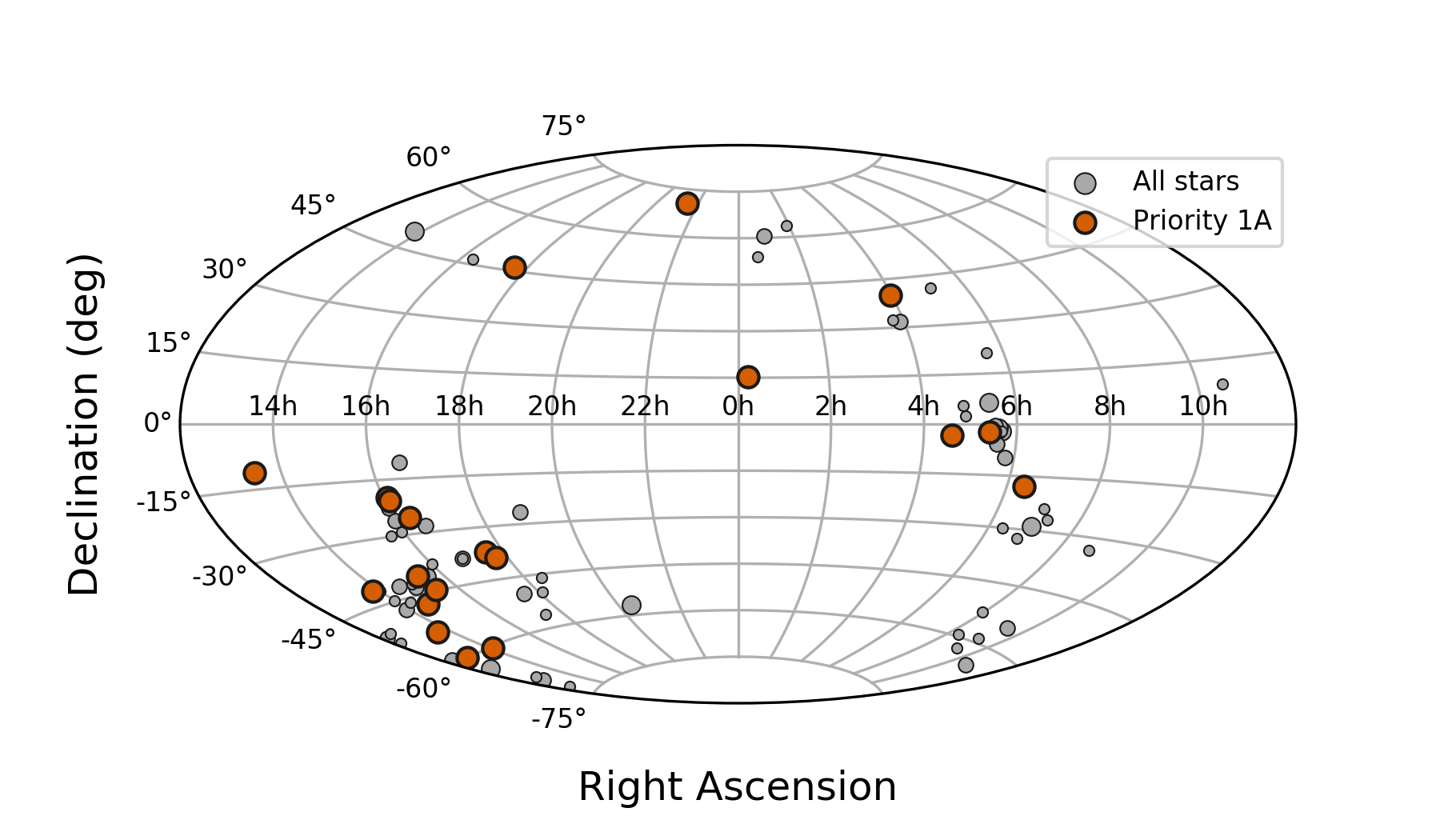} 
\caption{Left: SNR curves for simulated B0~V stars of different $V$ mag brightnesses for conservative instrument performance allocation and best case performance scenario using exposure times of 120 and 900~s. The SNR is estimated at a wavelength of 5000~\AA{} assuming $R = 55000$. Right: Sky distribution of high-priority CubeSpec targets from \cite[Bowman et al. (2022)]{Bowman2022}.}
\label{figure: targets}
\end{center}
\end{figure}

The compact design and science requirements of CubeSpec mean that bright (i.e. $V < 4$~mag) pulsating early-type stars are optimal targets to achieve its asteroseismic science case (see Fig.~\ref{figure: targets}). CubeSpec targets $\beta$~Cep stars, which have pulsation periods between $2 < P < 8$~hr and masses in the approximate range of $6 < M < 30$~M$_{\odot}$ \cite[(Bowman 2020)]{Bowman2020}. A search for pulsating stars with $V < 4$~mag and known spectral types between O8 and B3 (including dwarfs, giants and supergiants) was performed a priori using previous literature studies of photometric and spectroscopic variability and new TESS light curves \cite[(Bowman et al. 2022)]{Bowman2022}. In total, 23 $\beta$~Cep stars were identified from a sample of 90 stars, with the highest priority targets being slow rotators with large amplitude and long period pulsations. Their sky location is shown in Fig.~\ref{figure: targets}. Therefore, ample pulsating massive stars exist to be targeted by CubeSpec and empower asteroseismology of massive stars.

\end{document}